\documentstyle[psfig]{aa}

\begin{document}
   
%
   \title{Environmental effects in galaxies}

   \subtitle{The Data{\thanks{Based on 
   observations at the European Southern Observatory at the 
   15m Swedish ESO Submillimetre telescope, SEST, and at the 
    the 1.52m telescope which is operated under the ESO-ON agreement.}}}

   \author{Duilia F. de Mello
          \inst{1}
          Marcio A. G. Maia\inst{2}
	   \and
Tommy Wiklind\inst{1}
          }

   \offprints{D. de Mello}

   \institute{Onsala Space Observatory,
              43992 Onsala, Sweden\\
              email: duilia@oso.chalmers.se, tommy@oso.chalmers.se 
	     \and
      Observat\'orio Nacional, Rua Gal. Jos\'e Cristino 77, RJ 20921, Brazil\\
      email: maia@on.br}      

   \date{Received; accepted}

   \abstract{
   We present optical and millimetric data for 47 intermediate Hubble type 
spiral galaxies located either in dense environments or in the field. 
We compare correlations between global parameters, such as far-infrared luminosity, 
blue luminosity, and total molecular gas content, with other samples of galaxies, 
including normal galaxies, clusters and ultraluminous infrared galaxies. We find that 
overall our sample is a well defined subset of these other samples of galaxies.
}

\maketitle


\section{Introduction}

A longstanding issue in galaxy evolution is whether galaxies evolve 
according to a given set of initial conditions or whether the environment 
in which they reside is decisive for their evolution; i.e. whether 
galaxy evolution depends on nature or nurture. In order to search for
environmental effects in galaxies properties we have obtained optical and 
millimetric data for galaxies in dense regions of the Southern sky and in 
the field. In de Mello et al. (2001, hereafter Paper II) we present 
an extensive analysis of the data. The main results we found are:
intermediate type spirals in dense environments have on average less 
molecular gas per blue luminosity, lower current SFR, the same SFE 
and higher atomic gas fraction when compared with 
field galaxies. Although none of the above results stand out as a single strong 
diagnostic, given their statistical significance (see Table 3 of Paper II), taken together 
they suggest a trend for diminished gas content and star formation activity in galaxies in 
high density environments. We also found that SFR per blue luminosity 
increases linearly as the total amount of gas increases in LINERs. This result, based on a 
small sample, suggests that LINERs are powered by star formation rather than an AGN.
We refer to Paper II for a more detailed analysis of these results. 

In this paper we present the optical and millimetric data and is organized as 
follows. Section 2 describes the sample, Section 3 describes the optical data, 
Section 4 describes the millimetric data, Section 5 describes general properties, 
and a comparison with other samples, Section 6 presents a summary and conclusions.
A Database of optical and millimetric spectra together with digitized images
are shown in Appendix A (available at http://www.oso.chalmers.se/$\sim$duilia/env.html).


\section{Sample Selection}

\subsection{Previous samples}

Surveys of the molecular gas content in galaxies have in general been done
on samples which are far--infrared selected, or galaxies selected exclusively
for belonging to clusters or groups (often with a far--infrared selection
criteria on top; e.g. Casoli et al. 1991, Combes et al. 1994, Leon et al. 1998). A few exceptions exist in the literature. For example,
Sage (1993) presents the CO content of a distance limited sample of 65 
non-strongly interacting spiral galaxies, and Horellou et al. (1995) 
present a CO and HI survey of spiral and lenticular galaxies in the Fornax cluster, both 
based on samples selected without a far-infrared criterion. 

However, until now no survey of galaxies in different environments 
has included a rigorously selected control sample. For instance, the sample by Casoli et al. (1998) 
which contains a large sample of 582 objects is an important source of information concerning molecular 
gas in spiral galaxies. However, it was built by gathering data from various surveys and is very heterogeneous 
in terms of morphology and environment. It contains galaxies from several clusters as well as galaxies in the 
field.

\subsection{Dense Environment and Control Sample (HDS and CS)}

In view of these biases plagueing existing samples we have selected our sample 
from the catalog by Maia et al. (1994) which contains objects in low and high 
density areas of the Southern sky.
The selection of groups adopted by Maia et al. is similar to the methodology 
developed by Huchra \& Geller (1982) with the adaptations described by Maia et al. 
(1989). The catalog was drawn from the ESO/Uppsala Survey of the ESO(B) 
Atlas (Lauberts 1982) and used velocity information from the Southern Sky Redshift 
Survey (e.g., da Costa et al. 1989). The groups are defined to be formed by the accumulation of 
galaxy pairs with a member in common. 

\begin{itemize}

\item The high density sample ({\bf HDS}, hereafter) is formed by galaxies that are in groups of three or
more members. They have a density contrast
$\delta\rho/\rho \geq 500$. This is equivalent to densities larger than 18
galaxies/Mpc$^{3}$. All the selected objects have radial velocities
(after correction for Virgo infall) smaller than 8000 kms$^{-1}$.

\item The control sample ({\bf CS}, hereafter) is made up of galaxies which are not members
of any group and which are situated in regions with density contrast
$\delta\rho/\rho \leq 0.01$, i.e. less than 0.0004 galaxies/Mpc$^{3}$. 

\end{itemize}

\subsection{HDS versus Compact Groups and Poor Groups}

Although a group finding algorithm was used to generate the samples, the idea is 
not to identify groups (either loose or compact), but galaxies in high and low 
local density environments. The main difference between the HDS
and compact groups of galaxies is the isolation criterion which is imposed by the
groups selection (Hickson 1982, Coziol et al. 2000). 
The only 2 compact groups (HCG 21 and HCG 90) in the region searched 
by Maia et al. (1994) (b$^{II}$$\le$~--30$^{\circ}$, $\delta$$<$--17$^{\circ}$.5) 
have 3 galaxies of each group taking part of the HDS, but none of them take 
part in the present subsample analysis. 

The HDS should also not be confused with poor groups which are defined as systems with 
less than five bright galaxies but which can have 20-50 faint members (e.g, Zabludoff \& Mulchaey 1998, 
Willmer et al. 1999). Some galaxies in these poor groups are certainly 
part of the HDS, but since our selection includes only members with known redshift, 
the HDS will have only the brighter members which have measured redshift. The HDS 
and CS contain in total 151 and 179 galaxies, respectively.

\subsection{Our Subsample: Morphology selection}

Maia et al. (1994) have analysed the morphology distribution of the HDS and CS 
and concluded that the HDS has an excess of early-type galaxies compared to 
the CS. This is interpreted as an effect of the morphology-density 
relationship (Dressler 1980); i.e. a correlation between morphological types and
local density showing that the fraction of early-type galaxies increases as a
function of local galaxy density while the fraction of later types decreases (see
also Sanroma \& Salvador-Sol\'e 1990, Whitmore \& Gilmore 1991). Since there 
are galaxies of all morphologies in the HDS and in the CS, the main goal of our
work is to evaluate the effects of the environment in galaxies of the
{\it same morphological type} when compared with isolated galaxies. The ideal
survey would include all galaxies in the HDS and CS, however, due to large size of
the samples we have imposed such a selection which is fundamental in order 
to avoid any bias due to the well known correlation between morphology and physical 
properties of galaxies. Figures 2--4 of Roberts \& Haynes (1994) summarize clearly 
how morphology is correlated to fundamental properties of galaxies such as, blue
luminosity, far infrared lumninosity, total mass, and neutral hydrogen mass. One
of their conclusions is that, although the scatter is large, Sa-Sc have near
constant molecular gas normalized either by the blue luminosity or by the 
total mass. They also pointed out that later type spirals have less molecular gas
and suggest that this could also be due to the CO to H$_{2}$ conversion factor which 
would depend on morphology. Therefore, in order to have an homogeneous sample, we selected mostly intermediate 
spiral galaxies; i.e. Sb, Sbc, and Sc, avoiding Sa and Sd galaxies. In this work we 
present the analysis of the optical and millimetric data of a subsample of 47 spiral 
galaxies, 22 in the HDS and 25 in the CS, with velocities smaller than 5500 kms$^{-1}$.

Table~\ref{sample} lists information taken from the NASA/IPAC Extragalactic Database 
(NED) on each galaxy as follows. Column (1): designation
in the ESO-Uppsala catalog (LV89); column (2): designation in other
catalogs; column (3): right ascension ($^{\rm h\ m\ s}$) and declination ($^{\circ}$ $'$ $''$) for 
J2000; column (4): type of sample (control sample=CS and high density sample=HDS) and
morphological type 
(Lauberts \& Valentijn 1989, hereafter LV89) 1=Sa, 2=Sa-b, 3=Sb, 4=Sb-c, 5=S..., 6=Sc, Sc-d, 7=S../Irr, 8=Sd; 
column (5): morphological type from 
The Third Reference Catalogue of Bright Galaxies (RC3; de Vaucouleurs et al. 1991); column (6): number of galaxies
in the same group (Maia et al. 1989); column (7): mean pairwise separation in Mpc (Maia
et al. 1989); column (8): B$_{\rm T}$ magnitude from RC3; column (9): IRAS 60 $\mu$m flux in
Jy (Moshir et al. 1990) , and column (10): IRAS 100 $\mu$m flux in Jy (Moshir et al. 1990).

\begin{table*}
\caption[]{Observed Sample}
\label{sample}
\begin{tabular}{lccllccccc}
\hline
\multicolumn{1}{c}{ESO-LV}   & 
\multicolumn{1}{c}{Other} & 
\multicolumn{1}{c}{Coord.}& 
\multicolumn{1}{c}{Sample \&} & 
\multicolumn{1}{c}{Morph.} & 
\multicolumn{1}{c}{N$_{\rm g}$} & 
\multicolumn{1}{c}{r$_{\rm p}$} & 
\multicolumn{1}{c}{B$_{\rm T}$} &
\multicolumn{1}{c}{F$_{60\mu m}$} &
\multicolumn{1}{c}{F$_{100\mu m}$} \\
\multicolumn{1}{c}{Name} &
\multicolumn{1}{c}{Name} &
\multicolumn{1}{c}{J2000} &
\multicolumn{1}{c}{Morph.} &
\multicolumn{1}{c}{RC3} &
\multicolumn{1}{c}{} &
\multicolumn{1}{c}{Mpc} &
\multicolumn{1}{c}{} &
\multicolumn{1}{c}{Jy} &
\multicolumn{1}{c}{Jy} \\
\multicolumn{1}{c}{(1)} &
\multicolumn{1}{c}{(2)} &
\multicolumn{1}{c}{(3)} &
\multicolumn{1}{c}{(4)} &
\multicolumn{1}{c}{(5)} &
\multicolumn{1}{c}{(6)} &
\multicolumn{1}{c}{(7)} &
\multicolumn{1}{c}{(8)} &
\multicolumn{1}{c}{(9)} &
\multicolumn{1}{c}{(10)} \\
\hline
5390050 &	& 00 17 10.1 -19 18 00 & CS 5	&  SAB(rs)c?   &    &	   &  13.53 &	 0.977 &   2.972\\
3500140 & N101  & 00 23 54.6 -32 32 09 & CS  6  &  SAB(rs)c    &    &	   &  13.37 &	 0.549 &   1.754\\
3520530 & N491  & 01 21 20.3 -34 03 48 & HDS 3  &  SB(rs)b:    &  3 & 0.18 &  13.21 &	 2.843 &   8.632 \\
2960380 &	& 01 32 27.4 -38 40 40 & CS  4  &  SAB(rs)c    &    &	   &  13.99 &	 0.516 &   1.779\\
4780060 &	& 02 09 19.1 -23 24 54 & CS 4	&  Sbc         &    &	   &  13.22 &	 3.543 &   9.112\\
5450100 & N907  & 02 23 01.7 -20 42 43 & HDS 5  &  SBdm? sp    &  5 & 0.40 &  13.21 &	 2.649 &   5.625\\
5450110 & N908  & 02 23 04.8 -21 14 03 & HDS 5    & SA(s)c	&    &     &  10.83 &	14.770 &  43.670\\
3550260 &	& 02 32 17.5 -35 01 50 & CS  4  &  SB(s)bc:    &    &	   &  13.80 &	 0.482 &   1.588 \\
3550300 &	& 02 37 36.4 -32 55 28 & CS  4  &  SB(rs:)bc:  &    &	   &  13.59 &	 0.881 &   3.137 \\
0310050 &	& 02 58 06.0 -74 27 24 & CS  3.5 & SAB(rs)bc   &    &	   &  14.07 &	 1.043 &   3.887\\
3570190 & N1310 & 03 21 03.7 -37 05 58 & HDS 5  &  SB(rs)cd    & 55 & 0.82 &  12.55 &	 0.881 &   3.345 \\
5480070 & N1325 & 03 24 25.6 -21 32 35 & HDS 3.5 & SA(s)bc     &  7 & 0.94 &  12.22 &	 0.631 &   3.211\\
5480310 & N1353 & 03 32 03.0 -20 49 04 & HDS 3   & SA(rs)bc    &  7 & 0.94 &  12.40 &	 2.420 &   8.786\\
5480380 & I1953 & 03 33 41.7 -21 28 45 & HDS 6   & SB(rs)d     &  7 & 0.94 &  12.24 &	 8.470 &  11.128\\
4190030 &	& 03 42 11.2 -27 51 47 & CS 4	& (R')SAB(rs)c &    &	   &  13.60 &	 1.334 &   3.361\\
4820430 & N1459 & 03 46 58.0 -25 31 11 & CS 4	 & SB(s)bc?    &    &	   &  13.62 &	 0.572 &   2.657\\
4200030 &	& 04 07 45.8 -29 51 30 & CS 5	&  SA(rs)bc    &    &	   &  13.52 &	 0.704 &   2.172\\
2010220 &	& 04 08 59.3 -48 43 42 & CS  5   & Sbc         &    &	   &  14.73 &	 0.356 &   1.466\\
1570050 & N1536 & 04 10 59.9 -56 28 48 & HDS 5.5 & SB(s)c pec: & 46 & 1.30 &  13.15 &	 0.475 &   1.649\\
4840250 & N1591 & 04 29 30.7 -26 42 44 & CS 2	&  SB(r)ab pec &    &	   &  13.77 &	 1.929 &   5.001\\
1190060 & N1688 & 04 48 23.8 -59 47 59 & HDS 7.5 & SB(rs)dm    & 14 & 0.85 &  12.57 &	 2.683 &   6.677\\
1190190 & N1703 & 04 52 51.9 -59 44 33 & HDS 5   & SA(s)c      & 14 & 0.85 &  11.90 &	 2.122 &   7.723\\
3050140 &	& 05 12 34.1 -39 51 36 & CS  5  &  SB(s)c      &    &	   &  14.13 &	 0.378 &   0.982\\
2030180 & N1803 & 05 05 26.6 -49 34 05 & CS  4   & Sbc:        &    &	   &  13.38 &	 0.277 &   0.715\\
1420500 & I4901 & 19 54 23.1 -58 42 50 & CS  5   & SAB(r)c     &    &	   &  12.29 &	 1.778 &   6.518\\
2340160 &	& 20 23 25.1 -50 32 43 & HDS 5   & SAB(s)bc pec&  4 & 0.68 &  14.56 &	 3.069 &   7.875\\
2850080 & N6902 & 20 24 27.7 -43 39 09 & HDS 4  &  SA(r)b      &  4 & 0.31 &  11.64 &	 0.826 &   3.924\\
1060120 & I5038 & 20 46 51.2 -65 01 00 & CS  6   & (R':)SB(s)bc&    &	   &  14.13 &	 0.723 &   2.460\\
2350550 &	& 21 05 55.4 -48 12 23 & HDS 5   & (R')SAB(rs)bc& 9 & 1.00 &  12.70 &	 0.461 &   2.840 \\
2350570 &	& 21 06 21.8 -48 10 14 & HDS 4   & Sbc: sp     &  9 & 1.00 &  14.45 &	 0.461 &   3.368\\
2860820 &	& 21 15 45.4 -42 25 33 & HDS 5  &  SAB(s)c     &  3 & 0.20 &  14.51 &	 0.337 &   1.032\\
2370020 & N7124 & 21 48 05.7 -50 33 51 & CS  4.5 & SB(rs)c     &    &	   &  13.10 &	 0.791 &   3.411\\
1890070 & N7140 & 21 52 15.3 -55 34 10 & CS  4   & (R'$_{2}$)SB(rs)b &  &   &  12.20 &	 2.183 &   5.886 \\
2880260 & N7162 & 21 59 39.0 -43 18 12 & HDS 5  & (R')SA(r)bc  &  4 & 0.20 &  13.29 &	 0.484 &   1.656\\
5320090 & N7167 & 22 00 30.9 -24 38 00 & CS 5	&  SB(s)c:     &    &	   &  13.22 &	 1.314 &   3.588\\
6010040 &	& 22 01 30.4 -22 04 15 & CS 4.6  & SB(s)c:     &    &	   &  14.58 &	 0.227 &   0.877\\
1080130 & N7191 & 22 06 51.3 -64 38 03 & HDS 3.5 & SAB(rs)c    &  5 & 0.48 &  13.80 &	 0.570 &   2.061\\
1080200 & I5176 & 22 11 55.0 -66 50 46 & CS  3.9 & SAB(s)bc?sp &    &	   &  13.54 &	 3.031 &   11.21\\
1460090 & N7205 & 22 08 34.4 -57 26 33 & CS  5   & SA(s)bc     &    &	   &  11.55 &	 8.861 &  25.960\\
4050180 & N7267 & 22 24 21.6 -33 41 38 & CS  1  & (R'$_{1}$)SB(rs)a & &    &  12.91 &	 2.081 &   4.930\\
4060250 & N7418 & 22 56 36.0 -37 01 47 & HDS 5  &  SAB(rs)cd   & 32 & 1.31 &  11.66 &	 4.344 &  15.010\\
4060330 & I5270 & 22 57 54.7 -35 51 30 & HDS 6  &  SB(rs)c     & 32 & 1.31 &  13.00 &	 3.076 &   8.398\\
4070140 &	& 23 17 39.7 -34 47 24 & CS 5	&  SB(s)c?     &    &	   &  13.48 &	 0.987 &   2.766\\
3470340 & N7599 & 23 19 21.1 -42 15 20 & HDS 3  &  SB(s)c      & 32 & 1.31 &  12.08 &	 5.408 &  21.750\\
2400110 &	& 23 37 49.7 -47 43 42 & HDS 4.8 & Sb	       &  3 & 0.18 &  13.20 &	 0.956 &   5.612\\
2400130 &	& 23 39 26.9 -47 46 27 & HDS 3  & (R'$_{1}$)SAB(rs)b&3&0.18&  13.99 &	 0.791 &   3.411\\
4710200 & N7755 & 23 45 51.8 -30 31 19 & CS 4.5 &  SB(r)bc     &    &	   &  12.56 &	 2.686 &   8.538\\
\hline	     
\end{tabular} 
\noindent Column(4): CS=control sample, HDS=high density sample; morphological types are: 
1=Sa, 2=Sa-b, 3=Sb, 4=Sb-c, 5=S..., 6=Sc, Sc-d, 7=S../Irr, 8=Sd. Column(6): N$_{\rm g}$ is the
number of companions from Maia et al. 1989. Column(7): r$_{p}$ is the mean 
pairwise separation from  Maia et al. 1989.
\end{table*}

\section{The Optical Data}

\begin{table*}
\caption[]{Parameters of Strong Emission Lines}
\label{emission}
\begin{tabular}{clcccccc}
\hline
\multicolumn{1}{c}{ESO-LV} & 
\multicolumn{1}{c}{Sample \&} &
\multicolumn{1}{c}{F(H$\beta$)$\times$10$^{-15}$} &
\multicolumn{1}{c}{F([OIII]5007)$\times$10$^{-15}$}& 
\multicolumn{1}{c}{F(H$\alpha$)$\times$10$^{-15}$} & 
\multicolumn{1}{c}{F([NII]6583)$\times$10$^{-15}$} & 
\multicolumn{1}{c}{EW(H$\alpha$)} &
\multicolumn{1}{c}{Type of}\\
\multicolumn{1}{c}{name} &
\multicolumn{1}{c}{Morph.} &
\multicolumn{1}{c}{ergs cm$^{-2}$ s$^{-1}$} &
\multicolumn{1}{c}{ergs cm$^{-2}$ s$^{-1}$} &
\multicolumn{1}{c}{ergs cm$^{-2}$ s$^{-1}$} &
\multicolumn{1}{c}{ergs cm$^{-2}$ s$^{-1}$} &
\multicolumn{1}{c}{\AA} &
\multicolumn{1}{c}{Activity}\\
\multicolumn{1}{c}{(1)} &
\multicolumn{1}{c}{(2)} &
\multicolumn{1}{c}{(3)} &
\multicolumn{1}{c}{(4)} &
\multicolumn{1}{c}{(5)} &
\multicolumn{1}{c}{(6)} &
\multicolumn{1}{c}{(7)$^{\dagger}$} &
\multicolumn{1}{c}{(8)} \\
\hline
0310050& CS  3.5 & 4.0 &  0.3 &15.0 &  8.2&  8.2 &  HII \\
1060120& CS  6   &10.2 &  1.5 &38.0 & 18.0& 12.4 &  HII \\
1080130& HDS 3.5 & 7.2 &  0.6 &41.0 & 15.0& 14.2 & HII \\
1190190& HDS 5   & 5.7 &  1.3 &12.0 &  6.7&  5.5 &  HII \\
1420500& CS  5   &10.0 &  2.8 &20.0 & 16.0&  3.0 & L \\
1460090& CS  5   &21.0 &  5.1 &60.0 & 31.0&  5.6 &  HII \\
1570050& HDS 5.5 &14.5 &  4.4 &53.0 & 17.0& 27.9 & HII \\
2010220& CS  5   &12.0 &  8.8 &45.8 & 13.1& 16.6 &  HII \\
2030180& CS  4   &45.4 & 23.3 &160  & 63.0& 29.2 &  HII \\
2340160& HDS 5   &11.6 &  8.0 &40.0 & 15.0& 29.6 & HII \\
2350550& HDS 5   & 3.0 &  2.4 & 8.7 & 17.0& & L    \\
2350570& HDS 4   & 2.1 &  2.7 & 5.9 & 11.0&  1.2 & L \\
2370020& CS  4.5 & 7.9 &  2.8 & 3.3 &  9.5&  0.9 & L \\
2400110& HDS 4.8 &12.0 &  0.9 &16.6 & 18.5&  1.9  & L \\
2850080& HDS 4   &14.6 &  4.2 & 9.0 & 17.0&  1.1 & L \\
2860820& HDS 5   & 8.0 &  0.9 &27.0 & 11.0& 14.6 & HII \\
2880260& HDS 5   &19.0 &  1.0 &49.0 & 32.0&  6.6 & L \\
2960380& CS  4   & 7.2 &  1.7 &26.0 & 10.0& 16.5 &  HII \\
3050140& CS  5   & 1.8 &  0.4 & 6.7 &  3.8&  8.7 &  HII \\
3500140& CS  6   & 8.7 &  2.6 &33.0 & 14.0& 13.3 &  HII \\
3550300& CS  4   &12.2 &  2.6 &22.2 & 18.0&  3.7 & L \\
3570190& HDS 5   &13.4 &  8.4 &50.0 & 22.0& 14.9 & HII \\
4060330& HDS 6   &43.0 & 21.0 &180  & 66.0& 27.4 & HII \\
4070140& CS 5	 &53.2 & 51.0 &190  & 58.0& 40.6 &  HII \\
4190030& CS 4	 & 5.1 &  1.6 &22.0 &  7.8& 24.2 &  HII \\
4200030& CS 5	 &15.2 &  4.7 &48.0 & 18.0& 13.3 &  HII \\
4710200& CS 4.5  &24.3 &  6.5 &98.0 & 47.0& 10.3 &  HII \\
4780060& CS 4	 &23.1 &  6.0 &63.0 & 26.0& 13.7 &  HII \\
4820430& CS 4	 & 6.9 &  2.6 &17.0 &  9.4&  6.7 &  HII \\
5320090& CS 5	 &16.8 &  7.3 &51.0 & 21.0& 13.4 &  HII \\
5390050& CS 5	 &16.7 &  9.2 &64.0 & 28.0& 20.0 & HII \\
5450100& HDS 5   &23.4 & 24.7 &95.0 & 26.0& 30.0 & HII \\
5480310& HDS 3   &16.0 &  1.0 &53.0 & 35.0&  4.2 & L \\
5480380& HDS 6   & 5.6 &  1.4 &24.0 &  9.6& 28.8 & HII \\
6010040& CS 4.6  & 8.6 &  1.0 &30.0 & 14.0&  6.6 & HII \\
\hline
\end{tabular}
\noindent Column (2): CS=control sample, HDS=high density sample; morphological types are: 
1=Sa, 2=Sa-b, 3=Sb, 4=Sb-c, 5=S..., 6=Sc, Sc-d, 7=S../Irr, 8=Sd.  Column (8): HII=activity typical of HII regions, L=activity typical of LINERs.

\noindent $^{\dagger}$ Only EW(H$\alpha$) is corrected for internal reddening.
\end{table*}

\begin{figure*}
\psfig{figure=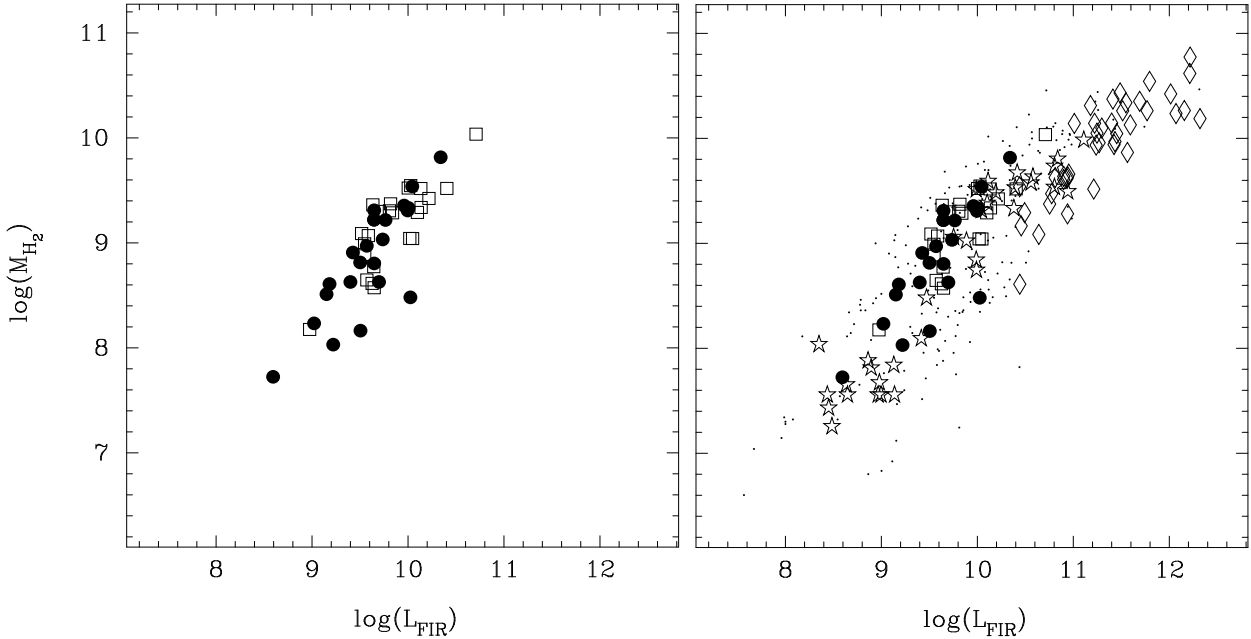,width=16.6cm,angle=270}
\caption[]{Left panel {\bf (a)}: Total molecular gas as a function of  FIR luminosity. 
The CS is marked by open squares and HDS by filled circles. Right panel {\bf (b)}: 
The same as in the left panel. Ultraluminous infrared galaxies (Sanders et al. 1991)
are marked by open diamonds, normal spiral galaxies from Young et al. (1989) and Braine et al. (1993)
are marked by dots, cluster galaxies (Coma and Fornax) from Casoli et al. (1991) and Horellou et al.
(1995) are marked by open stars. Luminosity is in L$_{\odot}$ and mass is in M$_{\odot}$.}
\label{MS1011f1}
\end{figure*}

Long slit spectra were obtained with the Boller \& Chivens Cassegrain 
spectrograph at the ESO 1.52m telescope (La Silla) during several runs in 1997 
and 1998 as part of a key project during Brazilian time. We used the Loral/Lesser 
CCD (No. 39) with 2048 $\times$ 2048 pixels (1 pixel = 15 $\mu$m) and 
grating No. 27 which has 600 lines mm$^{-1}$ and gives a spectral coverage of 
3600--7500 \AA\ and dispersion of 1.7 \AA~pix$^{-1}$. 
The slit width was 3$''$ and positioned along the major axis of the galaxies  
which corresponds to galactic sizes of 250~pc for the closest galaxy in the sample and to 
1~kpc for the most distant galaxy in the sample. 

Spectrophotometric standard stars were observed close to zenith several times 
during the night with a slit width of 5$''$. A He-Ar lamp was observed 
after every exposure and used for wavelength calibration. Typical exposure 
times were 2 $\times$ 20--30 minutes for galaxies and 5--10 minutes 
for stars. 

Standard data reduction, including bias and flat-field correction, 
was performed using IRAF. 
One dimensional spectra were extracted of each galaxy integrated along the slit length.
We corrected for Galactic extinction using Cardelli et al. (1989) extinction curve and
E(B-V) from NED. 
All spectra are flux calibrated and corrected for Doppler shift which was 
calculated using a cross-correlation technique.

Starlight subtraction was particularly critical in weak lines such as
H$\beta$. The starlight contribution was removed using the technique of McCall, Rybski \&
Shields (1985, see also Storchi-Bergmann, Calzetti \& Kinney 1994). 
Taking into account that in the typical stellar population the equivalent
width of H$\beta$ in absorption is of the order of 1.5\AA, we 
corrected for this effect by adding a factor of 1.5 times the continuum flux 
around H$\beta$ to the emission line flux. When no emission line was clearly 
visible we adopted a theoretical ratio, H$\alpha$/H$\beta$=2.86 (Ho et al. 1997). In this 
case, the value of H$\beta$ is an upper limit. Therefore, higher ratios of
H$\alpha$/H$\beta$ can also be expected. 
We have investigated whether a higher ratio would influence our results
by adopting ratios typical of AGNs (H$\alpha$/H$\beta$=3.1). We found no 
significant difference given the uncertainties in the continuum determination. 

We tested a second method of starlight subtraction using templates of old
stellar populations from Bica (1988). We subtracted our spectra from the
templates and then measured the fluxes. Both methods gave similar results 
given the accuracy of the measurements, dominated by the uncertainty in the 
continuum determination (Cid Fernandes et al. 1998). 

We measured the integrated fluxes and equivalent widths of the 
emission lines H$\beta$, [OIII]$\lambda$5007, [NII]$\lambda$6548, H$\alpha$, [NII]$\lambda$6583, 
[SII]$\lambda$6716,6731 for 35 galaxies with good signal-to-noise spectra.
Internal reddening was estimated from Cardelli et al. (1989) extinction curve and 
H$\alpha$/H$\beta$ ratios.
H$\alpha$ equivalent width was measured after internal reddening correction, following
the same procedure as in Ho et al. (1997). 

The type of activity was classified by measuring 
line-intensity ratios (log([OIII]~$\lambda$~5007/H$\beta$) and  
log([NII]~$\lambda$~6583/H$\alpha$)) and applying standard diagnostic diagrams 
(Baldwin et al. 1981, Veilleux \& Osterbrock 1987). In Paper II we show the diagnostic diagram
used to classify the type of activity.

Table~\ref{emission} lists the emission line 
parameters as follows. Column (1): designation in the ESO-Uppsala catalog  
(LV89); column (2): type of sample (control sample=CS and high density sample=HDS) and
morphological type 
(LV89) 1=Sa, 2=Sa-b, 3=Sb, 4=Sb-c, 5=S..., 6=Sc, Sc-d, 7=S../Irr, 8=Sd;
column (3): H$\beta$ flux; column (4): [OIII]$\lambda$5007
flux; column (5): H$\alpha$ flux; column (6):[NII]$\lambda$6583 flux; column (7): H$\alpha$
equivalent width in \AA, and column (8): type of activity (L=LINERS, HII=HII region). 

In Appendix A (available~at~the~following~webpage
http://www.oso.chalmers.se/$\sim$duilia/env.html)
we show the optical spectra of 35 galaxies of our sample. We also included in the
Appendix the CO spectra described below and images from The Digitized Sky 
Surveys~{\footnote{The Digitized Sky Surveys were produced 
at the Space Telescope Science Institute under U.S. Government grant 
NAG W-2166}} which allows direct inspection of the galaxies morphology.

\begin{table*}
\caption[]{CO Data}
\label{cotab}
\begin{tabular}{lllrrrrcc}
\hline	      
\multicolumn{1}{c}{ESO-LV} &
\multicolumn{1}{c}{Sample \&} &
\multicolumn{1}{c}{V$_{\rm CO}$} &
\multicolumn{1}{c}{$\Delta$V$_{\rm CO}$} &
\multicolumn{1}{c}{log L$_{\rm B}$} &
\multicolumn{1}{c}{L$_{\rm FIR}$$\times$$10^{9}$} &
\multicolumn{1}{c}{I$_{\rm CO(1-0)}$} &
\multicolumn{1}{c}{M$_{\rm H_2}$$\times$$10^{9}$} &
\multicolumn{1}{c}{I$_{\rm CO(2-1)}$} \\
\multicolumn{1}{c}{name} &
\multicolumn{1}{c}{Morph.}&
\multicolumn{1}{c}{kms$^{-1}$} &
\multicolumn{1}{c}{kms$^{-1}$} &
\multicolumn{1}{c}{L$_{\odot}$} &
\multicolumn{1}{c}{L$_{\odot}$} &
\multicolumn{1}{c}{K kms$^{-1}$} &
\multicolumn{1}{c}{M$_{\odot}$} &
\multicolumn{1}{c}{K kms$^{-1}$} \\
\multicolumn{1}{c}{(1)} &
\multicolumn{1}{c}{(2)} &
\multicolumn{1}{c}{(3)} &
\multicolumn{1}{c}{(4)} &
\multicolumn{1}{c}{(5)} &
\multicolumn{1}{c}{(6)} &
\multicolumn{1}{c}{(7)} &
\multicolumn{1}{c}{(8)} \\
\hline      
0310050 & CS  3.5 & 4714 & 287 &10.11 &13.60 $\pm$ 0.39 & 3.62 $\pm$ 0.26 & 3.30 $\pm$ 0.23& 3.64 $\pm$ 0.22\\
1060120 & CS  6   & 4154 & 180 & 9.97 & 6.90 $\pm$ 0.39 & 2.75 $\pm$ 0.28 & 1.93 $\pm$ 0.19& \\
1080130 & HDS 3.5 & 2941 & 135 & 9.78 & 2.67 $\pm$ 0.15 & 2.43 $\pm$ 0.21 & 0.81 $\pm$ 0.07& \\
1080200 & CS  3.9 & 1720 & 183 & 9.37 & 4.45 $\pm$ 0.20 & 6.28 $\pm$ 0.25 & 0.65 $\pm$ 0.03& 3.63 $\pm$0.18\\
1190060 & HDS 7.5 & 1256 & 43  & 9.48 & 1.66 $\pm$ 0.05 & 1.99 $\pm$ 0.14 & 0.11 $\pm$ 0.01& \\
1190190 & HDS 5   & 1527 & 33  & 9.94 & 2.52 $\pm$ 0.07 & 5.06 $\pm$ 0.19 & 0.42 $\pm$ 0.02& \\
1420500 & CS  5   & 2135 & 165 &10.10 & 4.40 $\pm$ 0.09 & 3.40 $\pm$ 0.25 & 0.59 $\pm$ 0.04& \\
1460090 & CS  5   & 1652 & 183 &10.13 &10.40 $\pm$ 0.34 &11.68 $\pm$ 0.66 & 1.10 $\pm$ 0.06& \\
1570050 & HDS 5.5 & 1311 & 40  & 9.30 & 0.39 $\pm$ 0.02 & 1.66 $\pm$ 0.06 & 0.10 $\pm$ 0.01& 0.88$\pm$ 0.09\\
1890070 & CS  4.0 & 3006 & 169 &10.44 & 9.15 $\pm$ 0.36 & 4.22 $\pm$ 0.26 & 1.48 $\pm$ 0.09& \\
2010220 & CS  5   & 3990 & 188 & 9.70 & 3.53 $\pm$ 0.23 & 1.50 $\pm$ 0.11 & 0.98 $\pm$ 0.07& 1.30 $\pm$0.11\\
2030180 & CS  4   & 4123 & 157 &10.27 &25.39 $\pm$ 1.17 & 4.66 $\pm$ 0.23 & 3.30 $\pm$ 0.16& 5.04 $\pm$0.14\\
2340160 & HDS 5   & 5218 & 10  &10.01 & 3.72 $\pm$ 0.50 & 0.82 $\pm$ 0.05 & 0.94 $\pm$ 0.06& 0.45 $\pm$0.12 \\
2350550 & HDS 5   & 5098 & 70  &10.73 & 9.92 $\pm$ 1.11 & 1.88 $\pm$ 0.11 & 2.04 $\pm$ 0.12& 1.34 $\pm$0.11 \\
2350570 & HDS 4   & 5069 & 248 &10.03 &11.08 $\pm$ 1.30 & 3.22 $\pm$ 0.12 & 3.45 $\pm$ 0.13& 3.56  $\pm$0.22\\
2370020 & CS  4.5 & 5214 & 236 &10.58 &13.72 $\pm$ 0.70 & 4.42 $\pm$ 0.16 & 4.90 $\pm$ 0.18& 1.96 $\pm$0.11\\
2400110 & HDS 4.8 & 2890 & 278 &10.00 & 5.81 $\pm$ 0.31 & 5.20 $\pm$ 0.17 & 1.65 $\pm$ 0.05& \\
2400130 & HDS 3  &  3284 & 50  & 9.80 & 5.17 $\pm$ 0.31 & 2.58 $\pm$ 0.15 & 1.08 $\pm$ 0.06& \\
2850080 & HDS 4  &  2838 & 132 &10.63 & 4.45 $\pm$ 0.23 & 1.97 $\pm$ 0.21 & 0.64 $\pm$ 0.07& \\
2860820 & HDS 5  &  4958 & 134 & 9.98 & 4.42 $\pm$ 0.51 & 1.62 $\pm$ 0.09 & 1.66 $\pm$ 0.09& 1.68 $\pm$0.08\\
2880260 & HDS 5  &  2383 & 80  & 9.79 & 1.42 $\pm$ 0.10 & 1.51 $\pm$ 0.11 & 0.32 $\pm$ 0.02& \\
2960380 & CS  4  &  3645 & 51  & 9.90 & 3.73 $\pm$ 0.33 & 0.84 $\pm$ 0.12 & 0.44 $\pm$ 0.06& \\
3050140 & CS  5  &  4761 & 450 &10.11 & 4.31 $\pm$ 0.55 & 2.38 $\pm$ 0.10 & 2.31 $\pm$ 0.10& 1.21 $\pm$0.08 \\
3470340 & HDS 3  &  1671 & 117 & 9.92 & 7.85 $\pm$ 0.79 &23.46$^\dagger$ $\pm$ 0.62 & 2.27 $\pm$ 0.06 & \\
3500140 & CS  6  &  3400 & 35  &10.09 & 3.30 $\pm$ 0.25 & 2.68 $\pm$ 0.09 & 1.23 $\pm$ 0.04& \\
3520530 & HDS 3  &  3874 & 260 &10.27 &21.89 $\pm$ 0.93 &10.85 $\pm$ 0.52 & 6.55 $\pm$ 0.32& \\
3550260 & CS  4  &  1985 & 105 & 9.42 & 0.95 $\pm$ 0.07 & 1.02 $\pm$ 0.13 & 0.15 $\pm$ 0.02& \\
3550300 & CS  4  &  4448 & 336 &10.25 &10.05 $\pm$ 0.43 & 4.08 $\pm$ 0.44 & 3.33 $\pm$ 0.36& \\
3570190 & HDS 5  &  1789 & 66  & 9.83 & 1.52 $\pm$ 0.06 & 3.40 $\pm$ 0.28 & 0.41 $\pm$ 0.03& \\
4050180 & CS  1  &  3375 & 124 &10.27 &10.67 $\pm$ 0.68 & 7.70 $\pm$ 0.37 & 3.52 $\pm$ 0.17& \\
4060250 & HDS 5  &  1470 & 83  & 9.98 & 4.42 $\pm$ 0.20 &27.44$^\ddagger$ $\pm$ 0.74 & 2.04 $\pm$ 0.05 & \\
4060330 & HDS 6  &  1922 & 110 & 9.71 & 5.01 $\pm$ 0.21 & 3.15 $\pm$ 0.17 & 0.42 $\pm$ 0.02& \\
4070140 & CS 5   &  2761 & 129 & 9.85 & 3.54 $\pm$ 0.23 & 2.64 $\pm$ 0.14 & 0.78 $\pm$ 0.04& \\
4190030 & CS 4   &  4146 & 83  &10.20 &11.19 $\pm$ 0.36 & 1.52 $\pm$ 0.13 & 1.10 $\pm$ 0.09& \\
4200030 & CS 5   &  4093 & 163 &10.22 & 6.41 $\pm$ 0.41 & 2.86 $\pm$ 0.20 & 2.02 $\pm$ 0.14& \\
4710200 & CS 4.5 &  3017 & 160 &10.30 &12.49 $\pm$ 0.53 & 5.46 $\pm$ 0.36 & 1.95 $\pm$ 0.13& \\
4780060 & CS 4   &  5401 & 164 &10.58 &51.12 $\pm$ 2.91 & 8.79 $\pm$ 0.37 & 10.86 $\pm$ 0.46& 15.16$\pm$0.33\\
4820430 & CS 4    & 4073 & 85  &10.17 & 6.57 $\pm$ 0.33 & 3.36 $\pm$ 0.30 & 2.35 $\pm$ 0.21& \\
4840250 & CS 2   &  4128 & 191 &10.13 &16.54 $\pm$ 0.64 & 3.63 $\pm$ 0.31 & 2.65 $\pm$ 0.22& \\
5320090 & CS 5   &  2582 & 83  & 9.91 & 4.22 $\pm$ 0.20 & 1.54 $\pm$ 0.17 & 0.41 $\pm$ 0.05& \\
5390050 & CS 5   &  3158 & 256 & 9.98 & 5.03 $\pm$ 0.29 & 4.95 $\pm$ 0.33 & 1.99 $\pm$ 0.13& \\
5450100 & HDS 5  &  1715 & 21  & 9.55 & 3.21 $\pm$ 0.14 & 1.29 $\pm$ 0.09 & 0.15 $\pm$ 0.01& 2.67 $\pm$ 0.13 \\
5450110 & HDS 5  &  1456 & 168 &10.35 &14.78 $\pm$ 0.72 &27.01 $\pm$ 0.92 & 2.15 $\pm$ 0.07& \\
5480070 & HDS 3.5 & 1557 & 17  & 9.87 & 1.05 $\pm$ 0.05 & 1.79 $\pm$ 0.11 & 0.17 $\pm$ 0.01& \\
5480310 & HDS 3   & 1531 & 108 & 9.79 & 3.17 $\pm$ 0.13 & 6.99 $\pm$ 0.32 & 0.65 $\pm$ 0.03& \\
5480380 & HDS 6   & 1874 & 86  &10.03 &10.56 $\pm$ 0.31 & 2.14 $\pm$ 0.15 & 0.30 $\pm$ 0.02& \\
6010040 & CS 4.6  & 5219 & 103 &10.01 & 3.85 $\pm$ 0.56 & 1.01 $\pm$ 0.08 & 1.17 $\pm$ 0.09& 0.78 $\pm$0.04\\
\hline
\end{tabular}

\noindent $^\dagger$ added CO(1~--~0) intensities of 5 points (map)
 
\noindent $^\ddagger$ added CO(1~--~0) intenstities of 7 points (map)

\noindent Column (2): CS=control sample, HDS=high density sample; morphological types are: 
1=Sa, 2=Sa-b, 3=Sb, 4=Sb-c, 5=S..., 6=Sc, Sc-d, 7=S../Irr, 8=Sd. 
\end{table*}

\section{The CO Data}

\begin{figure*}
\psfig{figure=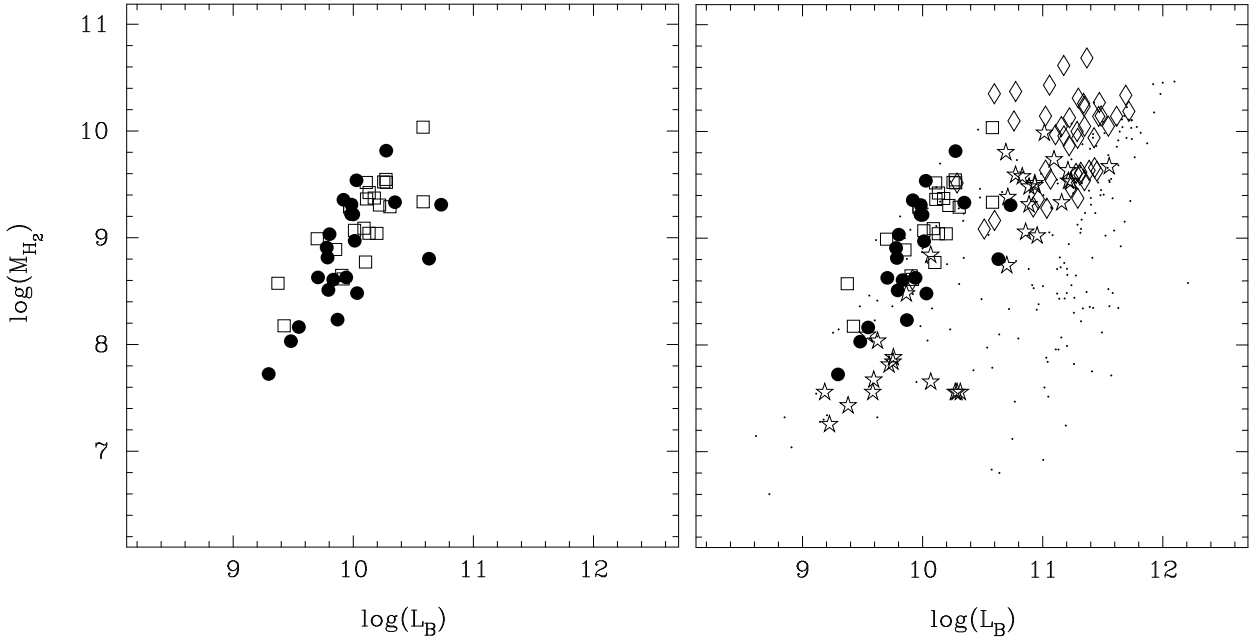,width=16.6cm,angle=270}
\caption[]{Left panel {\bf (a)}: Total molecular gas as a function of blue luminosity.
Right panel {\bf (b)}: The same as in the left panel. Additional samples of ultraluminous infrared
galaxies, galaxies in clusters, and spiral galaxies are included. Symbols are the same as in 
Fig.~\ref{MS1011f1}. Luminosity is in L$_{\odot}$ and mass is in M$_{\odot}$.}
\label{MS1011f2}
\end{figure*}

Millimetric observations were carried out at the Swedish-ESO (SEST) 15m
radiotelescope at La Silla in October 1996 and September 1998 during good weather 
conditions. In the first run we used the SESIS 100 receiver with a 1 GHz bandwidth 
at 115 GHz ($^{12}$CO (1~--~0)). Typical system temperatures were $\sim$ 250 K 
(in the T$_{\rm A}^{*}$ scale) at the elevation of the sources and 
typical zenith opacities between 0.1-0.2. During the second run we used the IRAM 115 
and IRAM 230 receivers with 500 MHz and 1 GHz bandwith, at 115 GHz ($^{12}$CO (1~--~0)) 
and 230 GHz ($^{12}$CO (2~--~1)), respectively. The half power beamwidth of the SEST 
at 115 GHz is 45$''$ and 23$''$ at 230 GHz.

All galaxies were observed at the central optical coordinate. Integration times 
were 2--3 hours depending on the signal-to-noise achieved. The pointing was regularly checked on
nearby SiO masers. The pointing uncertainties were of the order of 5$''$.
CO emission was detected in 47 galaxies and had low signal-to-noise 
detection in only 5 galaxies, ESO-LV1080110 (HDS), ESO-LV1880170 (CS), ESO-LV2850050 (HDS), 
ESO-LV3550300 (CS), and ESO-LV6050070 (CS). We have not included these galaxies in our analysis. 

Two galaxies, ESO-LV3470340 and ESO-LV4060250, were considerably larger than the SEST beam and were 
observed in 5 and 7 positions, respectively, spaced by half of a beamwidth (23$''$). 
In Appendix A (available at http://www.oso.chalmers.se/$\sim$duilia/env.html) we show each position along the major axis of the galaxy
and give their spectra. We have added the intensities at each position in order to obtain 
the total CO intensity of each galaxy. 

The CO spectra were reduced with the CLASS package (Forveille et al. 1990). 
We have binned the spectra with a boxcar function. Spectra 
were corrected for first order baseline in most of the cases or third order in a
few obvious cases where first order did not give a good fit to the data. CO intensities 
were calculated by using the main-beam efficiency, $\eta$$_{\rm mb}$, values of 0.7 and 
0.5 for 115 GHz and 230 GHz, respectively. We estimated the 1 $\sigma$ uncertainty 
in the integrated line intensity taking into account the channel--to--channel noise (rms), 
the width of the emission profile ($\Delta$V) and the number of channels (N) that the emission profile 
covers (error = rms$\times$$\Delta$V$\times$N$^{-1/2}$).

Table~\ref{cotab} lists the CO data as follows.
Column (1): designation in the ESO-Uppsala catalog  
(LV89); 
column (2): type of sample (control sample=CS and high density sample=HDS) and
morphological type 
(LV89) 1=Sa, 2=Sa-b, 3=Sb, 4=Sb-c, 5=S..., 6=Sc, Sc-d, 7=S../Irr, 8=Sd;
column (3): velocity derived from central CO (1~--~0) profiles in kms$^{-1}$; 
column (4): the width of the emission profile in kms$^{-1}$; 
column (5): blue luminosity in L$_{\odot}$ derived from B$_{\rm T}$ magnitude (errors in L$_{\rm B}$ are within
10\% when the magnitude estimates in the RC3 have errors of 0.1mag); 
column (6): Far-Infrared luminosity in L$_{\odot}$ calculated as described in the next Section;  
column (7): CO intensity in the line {\it J}=(1~--~0) in K kms$^{-1}$ and errors;  
column (8): H$_{\rm 2}$ masses and errors in  M$_{\odot}$ estimated from the velocity 
integrated CO (1~--~0) emission as described in the next Section, and 
column (9): CO intensity in the line {\it J}=(2~--~1) in K kms$^{-1}$.
Distances were corrected for the Virgocentric flow 
according to model 3.1 in Aaronson  et al. (1982). Hubble constant value of 
75 kms$^{-1}$Mpc$^{-1}$ was adopted in all calculations. 

Table~\ref{others} lists the CO intensity in the line {\it J}=(1~--~0) available in 
the literature for 7 galaxies (4 in the HDS and 3 in the CS). The differences between the fluxes 
we have measured and the ones
obtained previously are due to (i) different sizes of the beam (Elfhag et al. 1996), 
(ii) baseline adjustments (Combes et al. 1994, Andreani et al. 1995), or short integration 
time (Horellou \& Booth 1997).

\section{General Properties}

\begin{table}
\caption[]{CO data from the literature}
\label{others}
\begin{tabular}{llcl}
\hline	      
\multicolumn{1}{c}{ESO-LV} &
\multicolumn{1}{c}{Sample} &
\multicolumn{1}{c}{I$_{\rm CO(1-0)}$} &
\multicolumn{1}{c}{References}\\
\multicolumn{1}{c}{name} &
\multicolumn{1}{c}{}&
\multicolumn{1}{c}{K kms$^{-1}$} &
\multicolumn{1}{c}{}\\
\hline	      
1060120 & CS  & 2.2 & Combes et al. 1994$^\dagger$ \\
1570050 & HDS & $<$1.2 & Horellou \& Booth 1997$^\dagger$ \\
3570190 & HDS & $<$0.6 & Horellou \& Booth 1997$^\dagger$ \\
4780060 & CS  & 5.4$\pm$1.8 & Andreani et al. 1995$^\dagger$ \\
4840250 & CS  & 3.5$\pm$0.7 & Andreani et al. 1995$^\dagger$ \\
5450110 & HDS & 12.2$\pm$0.8 & Elfhag et al. 1996$^\ddagger$ \\
5480380 & HDS & 4.4 & Combes et al. 1994$^\dagger$ \\
\hline
\end{tabular}

\noindent $^\dagger$ using SEST, $^\ddagger$ using Onsala 20m
\end{table}

\begin{table*}
\caption[]{Average Values}
\label{averages}
\begin{tabular}{lrrrr}
\hline	      
\multicolumn{1}{c}{Sample} &
\multicolumn{1}{c}{log L$_{\rm B}$} &
\multicolumn{1}{c}{log L$_{\rm FIR}$} &
\multicolumn{1}{c}{log M$_{\rm H_2}$} &
\multicolumn{1}{c}{EW(H$\alpha$)$^\dagger$}\\
\multicolumn{1}{c}{} &
\multicolumn{1}{c}{L$_{\odot}$} &
\multicolumn{1}{c}{L$_{\odot}$} &
\multicolumn{1}{c}{M$_{\odot}$} &
\multicolumn{1}{c}{\AA} \\
\hline	      
HDS mean     &  9.94 $\pm$ 0.33 & 9.59 $\pm$ 0.40 & 8.86 $\pm$  0.39  & 15.9 $\pm$ 11.3 \\
HDS median   &  9.94 $\pm$ 0.12 & 9.65 $\pm$ 0.28 & 8.91 $\pm$  0.40  & 14.2 $\pm$ 11.2  \\
CS  mean     & 10.08 $\pm$ 0.29 & 9.85 $\pm$ 0.35 & 9.18 $\pm$  0.39  & 8.7  $\pm$ 3.4  \\
CS  median   & 10.11 $\pm$ 0.14 & 9.82 $\pm$ 0.21 & 9.29 $\pm$  0.22  & 8.2  $\pm$ 3.3   \\
\hline	  
\end{tabular}

\noindent $^\dagger$ Without LINERs. 

\end{table*}

\begin{figure*}
\psfig{figure=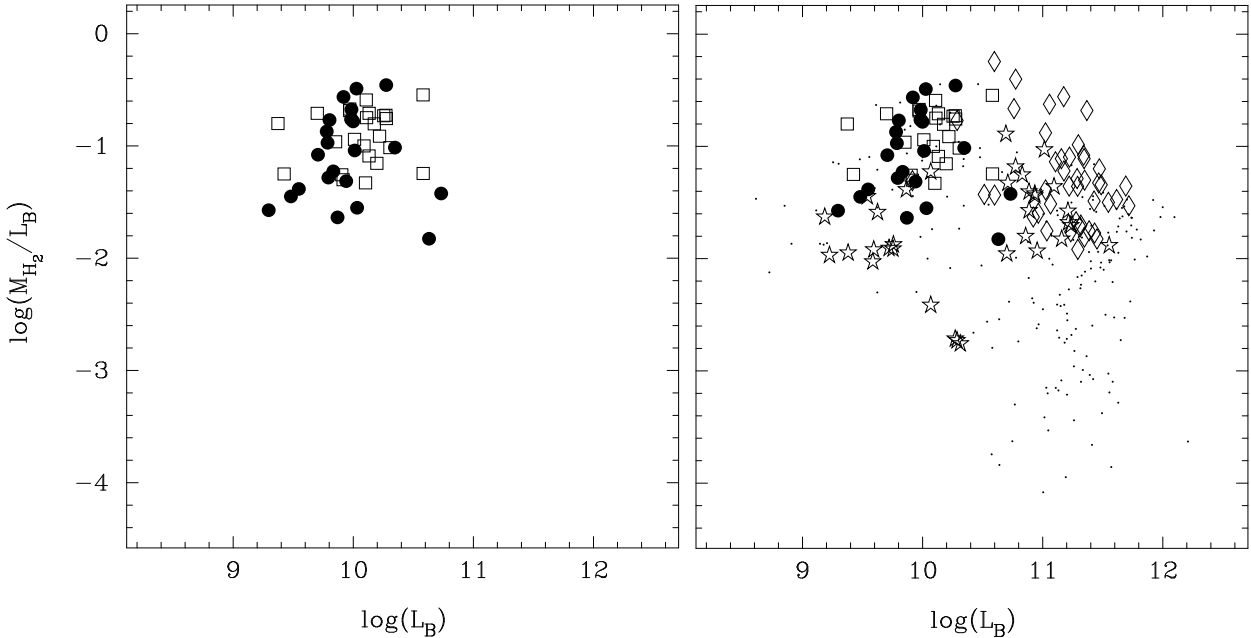,width=16.6cm,angle=270}
\caption[]{Left panel {\bf (a)}: total molecular gas normalized 
by the blue luminosity as a function of blue luminosity. Right panel {\bf (b)}: The same as in the left panel. Additional samples of ultraluminous infrared
galaxies, galaxies in clusters, and spiral galaxies are included.
Symbols are the same as 
in Fig.~\ref{MS1011f1}. Luminosity is in L$_{\odot}$ and mass is in M$_{\odot}$.}
\label{MS1011f3}
\end{figure*}

The FIR emission together with the molecular gas provide unique information in terms
of fuel and star formation. The FIR luminosity was calculated using the relation (Lonsdale \& Helou 1985) 

L$_{\rm FIR}$ = 5.9$\times$10$^{5}$D$^{2}$(2.58$\times$F$_{60}$+F$_{100}$)

\noindent where F$_{60}$ and F$_{100}$ are fluxes in Jy at 60 and 100 $\mu$m detected 
by IRAS and D is the distance in Mpc corrected for the Virgo infall.

H$_{\rm 2}$ masses were estimated from the velocity 
integrated CO (1~--~0) emission, using a 
N$_{\rm H_{2}}$/I$_{\rm CO}$ conversion ratio of 3$\times$10$^{20}$
cm$^{-2}$ (K kms$^{-1}$).

We are assuming that the conversion factor is the same in all galaxies in our
sample. This assumption is reasonable since our sample do not contain any
later-type systems (Sd, Sm, Ir) which, despite the ongoing star formation, 
show weak CO emission (e.g. Rubio et al. 1991). 

Average and median values of L$_{\rm B}$, L$_{\rm FIR}$, M$_{\rm H_{2}}$, and
H$\alpha$ equivalent width are presented in Table~\ref{averages}. 

Fig.~\ref{MS1011f1}a and Fig.~\ref{MS1011f2}a show the total 
amount of molecular gas as a function of FIR and blue luminosities. Fig.~\ref{MS1011f1}a 
confirms the known correlation between L$_{\rm FIR}$ and the H$_{\rm 2}$ masses 
(correlation coefficient= 0.80 and 0.84 for the HDS and CS, respectively). From Fig.~\ref{MS1011f2}a we verify that 
galaxies in the CS are on average more luminous than those in the HDS 
(a distance bias in our subsample). In order 
to eliminate this effect, CO intensities were normalized by the blue luminosity, L$_{\rm B}$, 
in the analysis presented in Paper II. Given our morphological selection criteria, we assumed that 
the mass/L$_{\rm B}$ ratio is approximately the same for our galaxies (Roberts \& Haynes 1994) 
and L$_{\rm B}$ is thus a measure of the total mass.

We have plotted the M$_{\rm H_{2}}$/L$_{\rm B}$ 
as a function of L$_{\rm B}$ (Fig.~\ref{MS1011f3}a) in order to compare whether the bias in blue luminosity present in our 
subsample may cause a bias in our analysis. The correlation found for HDS and CS is very
similar (correlation coefficient= -0.03 and 0.06 for the HDS and CS, respectively) suggesting no evident bias. 
We have compared our sample properties with samples observed by others, such as 
normal spiral galaxies (Young et al. 1989, Braine et al. 1993), the ultraluminous FIR galaxies 
(Sanders et al. 1991), and galaxies in the Coma and Fornax clusters (Casoli et al. 1991 
and Horellou et al. 1995). 
As it is shown in Fig.~\ref{MS1011f1}b, Fig.~\ref{MS1011f2}b, and Fig.~\ref{MS1011f3}b the 47 
spiral galaxies of our sample (HDS and CS) have correlations between global parameters which are similar 
to those in other samples. The ultraluminous FIR galaxies (Sanders \& Mirabel 1996), 
as expected, are overall brighter and more massive than our subsample. The other samples include 
spirals of all types which explains the large dispersion found in luminosities and masses.

\begin{figure*}
\psfig{figure=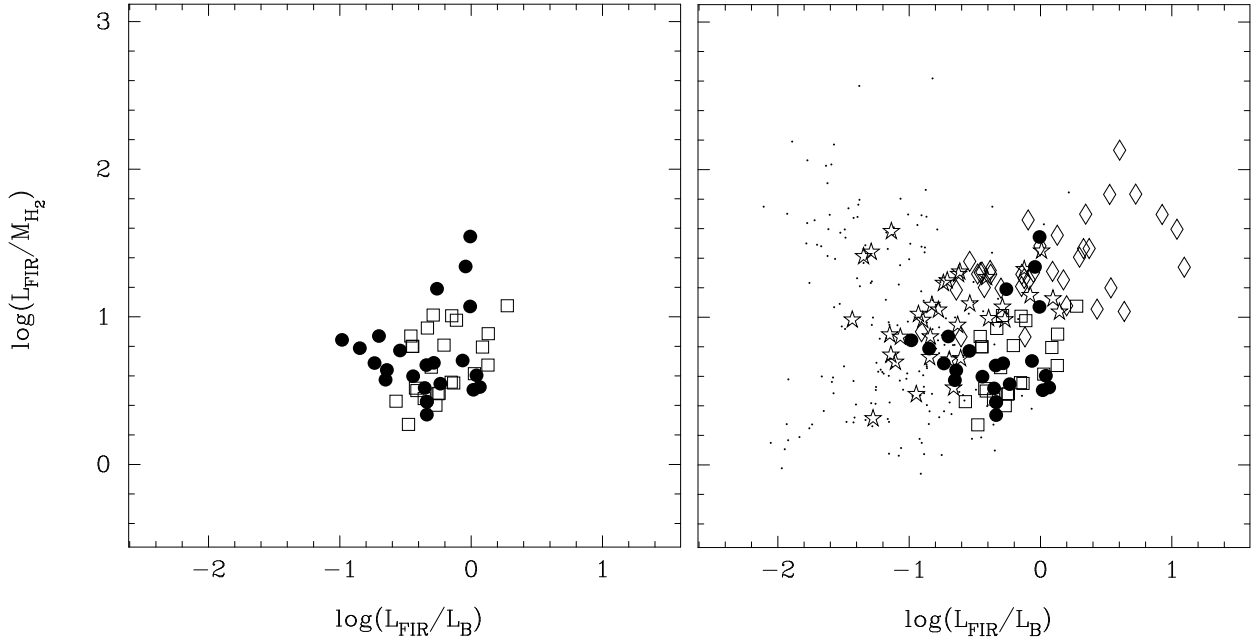,width=16.6cm,angle=270}
\caption[]{Left panel {\bf (a)}: FIR luminosity normalized by the total molecular gas as a function of the FIR luminosity normalized by the blue luminosity.
Right panel {\bf (b)}: The same as in the left panel. Additional samples of ultraluminous infrared
galaxies, galaxies in clusters, and spiral galaxies are included. Symbols are the same as in Fig.~\ref{MS1011f1}.
Luminosity is in L$_{\odot}$ and mass is in M$_{\odot}$.}
\label{MS1011f4}
\end{figure*}

As previously mentioned, only intermediate Hubble types (Sb, Sbc, and Sc) were selected in order to avoid any 
bias due to the correlation between general
properties and morphology. However, even in this sample the uncertainties in morphological 
classification should be taken into account when making any firm statement. 
Galaxies in dense environments can have their morphology distorted by tidal 
effects which makes them difficult to classify. One should refer to Appendix A 
(available at http://www.oso.chalmers.se/$\sim$duilia/env.html)
in order to visually check the morphology of each individual galaxy 
in more detail. We also refer to the detailed morphological classification taken from 
RC3 presented on Table~\ref{sample} which
gives a general idea on the complexity of the morphologies.

In Table~\ref{cotab} we give both the CO(1~--~0) and CO(2~--~1) integrated
line intensitites. In order to estimate the CO(2~--~1)/CO(1~--~0) intensity
ratios we need to convolve the CO(2~--~1) data to the same angular resolution
as the CO(1~--~0) data. Since we observed only a single position for most
galaxies, we can not do this. However, taking the values in Table~\ref{cotab} at
face value, the average CO(1~--~0) to CO(2~--~1) line intensity
ratio is 0.93$\pm$0.47. This is an upper limit to the line ratio. In the
case of a molecular gas distribution more extended than both the CO(1~--~0)
and CO(2~--~1) telescope beams (45$''$ and 23$''$, respectively), the
correction for different angular resolutions would be 1.0.
In the other extreme, with the CO emission originating in a point source,
the correction for different angular resolutions would be 0.25. Since our
telescope beam in almost all cases is large with respect to the optical extent
of the galaxies, and since the molecular gas is likely to be centrally
concentrated, the correction for different angular resolutions should be
$\sim$0.5.
Our average line ratio is thus $\sim 0.5 \pm 0.4$. This value is lower
than that found by Braine et al. (1993) of 0.89$\pm$0.34 for normal spiral
galaxies. The lower value is characteristic of optically thick and subthermally
excited molecular gas and most likely reflects the lower star formation
activity in our environmentally selected sample as opposed to far infrared
bright selected samples.

In Fig.~\ref{MS1011f4} we verify that the HDS and CS are also very similar to the 
galaxies in other samples in terms of SFE. We conclude that the intermediate type spirals
in the HDS and CS do not belong to a separate class of object but contain objects with properties 
similar to galaxies in clusters, nearby spiral galaxies and infrared luminous galaxies.

\section{Summary}

In this paper we present millimetric and optical data obtained in order to 
study environmental effects in galaxies. Our sample has 47 intermediate Hubble type
spirals in either dense environments or in the field. We compared general properties, 
such as far-infrared luminosity, blue luminosity, and total molecular gas content, 
to other samples of galaxies, such as ultraluminous infrared galaxies, clusters of
galaxies and spiral galaxies. We find that overall our sample has general properties
very similar to these other galaxies; i.e. they are not a separate class of objects.

\begin{acknowledgements}
The ON team of observers at the ESO1.52m, in particular Christopher Willmer for helping
with the data reduction. Henrique Schmitt for valuable suggestions regarding the stellar 
contamination. This research has made use of the NASA/IPAC Extragalactic Database (NED) which is
operated by the Jet Propulsion Laboratory, California Institute of Technology, under contract with the
National Aeronautics and Space Administration. The Digitized Sky Surveys were 
produced at the Space Telescope Science Institute under U.S. Government 
grant NAG W-2166. The images of these surveys are based on photographic 
data obtained using the Oschin Schmidt Telescope on Palomar Mountain and the 
UK Schmidt Telescope. 
D.F.M. was partially supported by CNPq Fellowship 301456/95-0, and the Swedish 
\emph{Vetenskapsr{\aa}det} project number F620-489/2000. M.A.G.M. was 
supported by CNPq grant 301366/86-1. T.W. was supported by \emph{Vetenskapsr{\aa}det} 
project number F1299/1999.
\end{acknowledgements}


\end{document}